\documentclass[preprint2]{aastex}

\begin{document}

\title{Nucleosynthesis of Nickel-56 from Gamma-Ray Burst Accretion Disks}

\author{R. Surman\altaffilmark{1}, G. C. McLaughlin\altaffilmark{2}, and N. Sabbatino\altaffilmark{1}}

\altaffiltext{1}{Department of Physics and Astronomy, Union College, Schenectady, NY 
12308}
\altaffiltext{2}{Department of Physics, North Carolina State University, 
Raleigh, NC  27695-8202}

\begin{abstract} 

We examine the prospects for producing Nickel-56 from black hole accretion disks, by examining a range of steady
state disk models.  We focus on relatively slowly accreting disks in the range of $\dot{M} = 0.05$ $M_\odot / {\rm
s}$ to $\dot{M} = 1$ $M_\odot / {\rm s}$, as are thought to be appropriate for the central engines of long-duration
gamma-ray bursts.  We find that significant amounts of Nickel-56 are produced over a wide range of parameter
space. We discuss the influence of entropy, outflow timescale and initial disk position on mass fraction of
Nickel-56 which is produced.  We keep careful track of the weak interactions to ensure reliable calculations of
the electron fraction, and discuss the role of the neutrinos.  

\end{abstract}


\section{Introduction}

Over the last decade and a half there has been a dramatic increase in data available for studying long duration
gamma-ray bursts (GRBs). Besides the gamma rays, afterglows in the x-ray, optical and radio are seen, and host
galaxies have been identified.  In a handful of cases ``supernova bumps'' are seen in the optical afterglows,
e.g.  GRB 980326 \citep{bloom99}.  Furthermore, spectra for Type Ic supernova have been identified in a few cases
\citep{galama98,hjorth03,kawabata03,matheson03,stanek03, malesani04,thomsen04}. These developments have lead to
the theoretical conjecture that all long duration gamma ray bursts are the result of the collapse of a massive
star; for a review see \citet{woosley06}. 

The temporal evolution of light curve spectra of traditional core collapse supernovae are driven by the decay of
Nickel-56 to Cobalt-56 and the subsequent decay of Cobalt-56 to Iron-56.  The half lives for these decays are
approximately six days and seventy-seven days respectively.  After these nuclei beta decay, their daughter nuclei
emit gamma rays which then thermalize in the ejecta. The energy from these decays is thus re-emitted in the optical
and observed as the supernova light curve.  Thus the intensity of a supernova bump should be directly influenced by
the amount of $^{56}$Ni produced in the object.  On the order of tenths of solar masses of Nickel are needed.  For
example, it has been estimated that around $0.5 {\rm M}_\odot$ of $^{56}$Ni is necessary to explain SN2003dh which
accompanied GRB030329 \citep{woosley03}.

The commonly discussed model for the type of supernova that produces a GRB is the collapsar model
\citep{woosley93,macfadyen99}. In this case the star has too much rotation to produce a standard core collapse
supernova.  A hot $\sim 10$ MeV disk surrounding a black hole is produced at the center.  Nickel-56 can be
produced in two different ways.  It can be produced in explosive burning, when a shock wave propagates out
through the star and pre-existing nuclei are then burned to $^{56}$Ni \citep{maeda03,fryer06,maeda09}. This is
the same mechanism as in traditional core collapse supernovae, but the elemental yields are different since the
explosion energy is larger.  However, Nickel-56 can also be produced in a disk wind. This is a primary process
where free neutrons and protons are ejected from the disk and combine to form nuclei \citep{sur05,pru04,sur06}. 

Any assessment of $^{56}$Ni production in GRBs must consider both mechanisms.  The latter mechanism is more
difficult to model, since it requires a careful accounting of the neutrinos in the disk. In this paper we extend
the preliminary studies of \citet{sur05,pru04,pru05,sur06}, and examine the production of Nickel-56 from the outflow
from a variety of disks.  We discuss not only the importance of the electron fraction in determining the
abundance yields, but also the role of entropy and outflow timescale.  Recent studies, e.g. \citet{maeda09} and
references therein, have used estimates of ejected $^{56}$Ni mass to constrain the GRB central engine.  This study
provides a reference for anyone considering Nickel-56 production in models of long duration gamma-ray bursts.
 
\section{Disk and Outflow models}

In order to study the nucleosynthesis in the hot outflows from GRB accretion disks, we first need to determine
the composition of and the neutrino fluxes emitted from the accretion disk itself.  We therefore begin with
existing hydrodynamical black hole accretion disk (AD-BH) models with parameters appropriate for long-duration
GRBs.  For this study we have chosen the one-dimensional, steady-state AD-BH models of \cite{che07} (CB).  The CB
models are fully relativistic and incorporate important microphysics effects, such as neutrino trapping and an
evolving neutron-to-proton ratio, neglected in other relativistic disk calculations.  Here we investigate a range
of steady-state disk models with mass accretion rates $0.05 \leq \dot{m} \leq 1$, where $\dot{m} = 1$ $M_\odot /
{\rm s}$. 

\begin{figure}[h]
\plotone{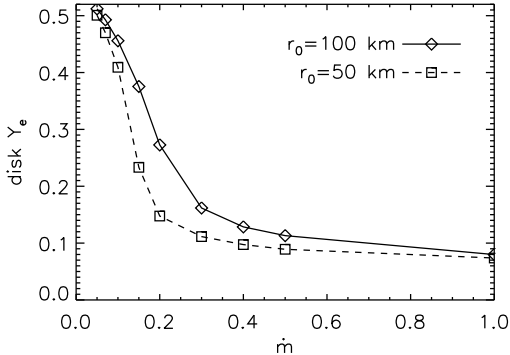}
\caption{Shows the electron fractions in the disk at radial distances of $r_{0}=50$ km (squares) and $r_{0}=100$ km
(diamonds) for AD-BH with mass accretion rates $\dot{m}$.\label{fig:diskye}}
\end{figure}

We calculate the initial nuclear composition of each disk as described in \cite{sur04}.  We find the disk
compositions we calculate as a post-processing step match well with the disk compositions reported in
\cite{che07}, which are calculated along with the disk hydrodynamics.  In the inner regions of all of the disks,
nuclei are dissociated into their constituent nucleons, and the ratio of neutrons to protons is subsequently set
by the weak interactions
\begin{eqnarray}
p + e^{-} & \rightleftharpoons & n + \nu_{e} \\
n + e^{+} & \rightleftharpoons & p + \bar{\nu}_{e}.
\end{eqnarray}
We describe the composition of the disks by the electron fraction $Y_{e}$, where $Y_{e}=1/(1+n/p)$ and $n$ and
$p$ are the number densities of neutrons and protons, respectively.  For disks with the lowest accretion rates of
those considered here, weak reaction rates are slow even in the inner regions of the disks.  In these cases the
disk composition remains roughly balanced between neutrons and protons and $Y_{e}$ remains close to 0.5.  For
disks with higher accretion rates, conditions in the inner region of the disks are hotter and denser, and weak
reactions are correspondingly faster.  For disks with $\dot{m}<1$, electron capture on protons, the forward
reaction of Eqn. 1, predominates, and the inner regions of the disks become quite neutron rich. 
Fig.~\ref{fig:diskye} shows the electron fractions in the disk for two radial distances, $r_{0}=50$ km and 100
km, from the black hole for the disk models used in this work.  A full discussion of the disk compositions of
AD-BH can be found in, for example, \cite{pru03,fuji03,sur04,che07}.  These disk electron fractions provide the
starting points for the calculations of the nucleosynthesis from the disks.

While calculating the composition of the disks, we additionally calculate the neutrino fluxes from the disk as in
\cite{sur04}.  Most of the AD-BHs considered in this work are optically thin to neutrinos and the emitted fluxes
are primarily due to electron and positron captures in the disk.  Since the rate of electron captures in the
disks far exceeds the rate of positron captures for all disks with $\dot{m}\geq 0.1$, the electron neutrino flux
emitted from the disk is significantly higher than the electron antineutrino flux.  This has important
consequences for the nucleosynthesis in the outflow.

\begin{figure}[h]
\epsscale{.7}
\plotone{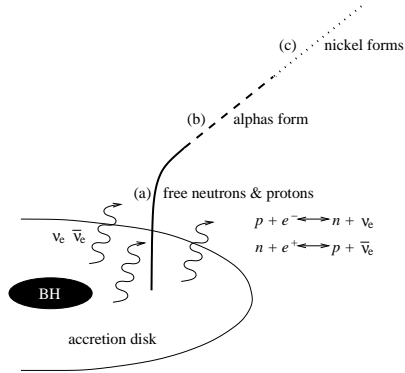}
\caption{Shows a schematic of an outflow trajectory, labeled with the three stages of outflow
nucleosynthesis.\label{fig:traj}}
\end{figure}

We follow the outflowing material as described in \cite{sur05}. We take the outflow to be adiabatic and the velocity $v$
of the outflowing material as a function of radial distance from the black hole $r$ to be
$v=v_{\infty}(1-r_{0}/r)^{\beta}$, with $v_{\infty}=10,000$ km/s, $r_{0}$ is the starting disk position, and $\beta$
controls the rate of the acceleration of the material.  We investigate a range of $\beta$, $0.2\leq \beta \leq 2.6$,
where smaller $\beta$ corresponds to more rapid initial acceleration, and two disk starting positions, $r_{0}=50$ km and
100 km.  We connect the evolution of the density to our velocity prescription by setting the mass loss rate $\dot{M}$ of
the outflow to be constant, first in cylindrical symmetry as the material lifts off vertically from the disk, and then
in spherical symmetry as the material expands radially away at later times.  We take the initial composition of the
material to be that of the disk at the starting radius $r_{0}$.  We then follow the nuclear recombination using a full
nuclear network code as described in \cite{sur06}.  A schematic of the outflow is shown in Fig.~\ref{fig:traj}.

While the entropy of the disk is quite low, several potential heating processes can increase the entropy of the material as
it lifts off the disk.  To make some estimate of what range in entropies is realistic for the wind, we consider the rate of
change of entropy due to just one potential heating source---neutrino captures in the outflow.  We start with the
approximate expressions for the neutrino heating and cooling rates from \cite{qia96} and calculate the change in entropy
per baryon from energy conservation, recast for a neutrino-driven wind in Eqn (26) of \cite{qia96} as
\begin{eqnarray}
v\frac{ds}{dr}&=&\frac{\dot{q}m_{N}}{T},
\end{eqnarray}
where $s$ is the entropy per baryon, $m_{N}$ is the mass of a nucleon, $T$ is the temperature, and $\dot{q}$ is the
specific net heating rate due to neutrino and antineutrino interactions.  When we apply this expression to our outflow
velocity and density profiles as described above, we find, for example, for an $\dot{m}=0.15$ disk, the change in
entropy per baryon in the outflow due to neutrino interactions ranges from $1-23k$, while for a $\dot{m}=1$ disk, the
range is $4-58k$.  We note that these estimates represent a lower limit, as additional processes (e.g., magnetic
heating) are expected to play an important role.  We therefore investigate outflow entropies up to somewhat larger
values, $10 \leq s/k \leq 100$.

\section{Production of Nickel-56}

\begin{figure}[h]
\epsscale{1}
\plotone{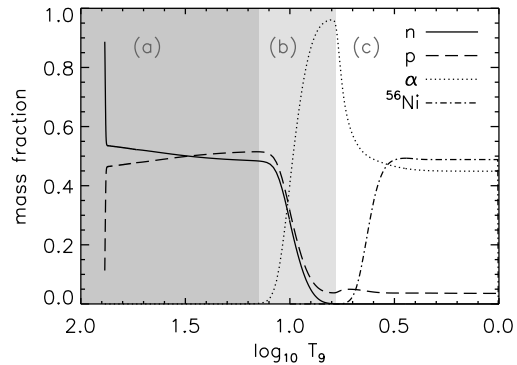}
\caption{Shows the mass fractions of neutrons (solid line), protons (dashed line), alphas (dotted line) and
$^{56}$Ni (dot-dashed line) as a function of decreasing temperature $T_{9}$ for an outflow trajectory with
parameters $r_{0}=100$ km, $s/k=30$, and $\beta=0.8$ from an AD-BH with $\dot{m}=0.5$.  The shaded areas refer to
the portions of the outflow trajectory where (a) the ratio of free protons to neutrons is set by weak
interactions (dark shaded region), (b) protons and neutrons combine to form alphas (light shaded region), and (c)
alphas combine to make $^{56}$Ni (unshaded region). \label{fig:nuc}}
\end{figure}

The element synthesis leading to the production of $^{56}$Ni for a sample AD-BH outflow trajectory is depicted in
Fig.~\ref{fig:nuc}, which shows the evolution of the mass fractions of neutrons, protons, alpha particles, and
$^{56}$Ni as a function of decreasing temperature $T_{9}$, where $T_{9}=T/10^{9}$ K and $T$ is the temperature in
the outflow in Kelvins.  In this example, the material is ejected from an AD-BH with $\dot{m}=0.5$ at a starting
radius of $r_{0}=100$ km; the initial composition is therefore quite neutron rich.  As the material lifts off the
disk, it transitions from the dense, low entropy, electron-degenerate disk conditions to the higher entropy
($s/k=30$ in this example), hotter, accelerated outflow conditions.  The higher entropy favors a balance of
electrons and positrons such that positron capture, the forward reaction of Eqn. 2, can compete with electron
capture, and the electron fraction quickly rises.  As the material moves away from the disk, it cools and
expands, and weak reaction rates slow.  Even so, the electron fraction continues to evolve as the positron
capture rate, and eventually the electron neutrino capture (the reverse reaction of Eqn. 1) rate, are slightly
faster than the electron capture rate.  The positron capture rate exceeds the electron capture rate because of
the neutron-proton mass difference, and eventually the electron neutrino capture rate exceeds both the electron
and positron capture rates as the latter drop much more rapidly ($\sim T^{-5}$) above the disk.  Therefore the
material gradually becomes proton-rich prior to nuclear reassembly, as seen in Fig.~\ref{fig:nuc} for $T_{9}>10$. 
This portion of the outflow trajectory, where the material is composed of free nucleons and their ratio is
adjusted by weak interactions, is labeled (a) in Figs.~\ref{fig:traj} and \ref{fig:nuc}.

In the first stage of nuclear reassembly, labeled (b) in Figs.~\ref{fig:traj} and \ref{fig:nuc}, protons and
neutrons combine into alpha particles ($^{4}$He nuclei).  Conversion into alphas is most efficient at
$Y_{e}=0.5$, where equal numbers of protons and neutrons are present.  In this example, the material is slightly
proton-rich, so all of the neutrons and a matching number of protons combine into alphas, and a small fraction of
excess protons are left over.  In the second stage of nuclear reassembly, labeled (c) in Figs.~\ref{fig:traj} and
\ref{fig:nuc}, alpha particles combine via the triple alpha process to produce $^{12}$C, and a series of alpha
captures on carbon culminate in the production of $^{56}$Ni. The triple-alpha process and the subsequent alpha
captures are highly temperature-dependent, and so the efficiency of the conversion of alphas to $^{56}$Ni depends
strongly on the evolution of the thermodynamic conditions in the outflow.  If the outflow is coasting at this
point such that the temperature drops slowly, most of the alphas will convert to nickel; in the example
considered here, however, this isn't quite the case, and a bit more than one half of the alphas end up in
$^{56}$Ni.

The production of $^{56}$Ni can therefore depend on a number of parameters: the mass accretion rate of the AD-BH,
the point in the AD-BH at which the outflow originates, the entropy per baryon and the acceleration of the
outflow, and the neutrino interactions in the outflow.  The influence of each of these parameters in turn is
described in sections \ref{entrop}-\ref{diskr} below.

\subsection{Variation with entropy} \label{entrop}

\begin{figure}[h]
\plotone{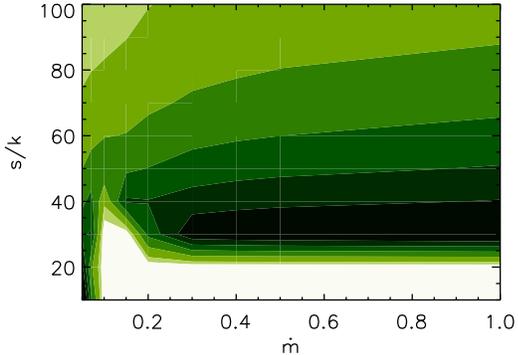}
\caption{Shows nickel production as a function of the entropy $s/k$ of the outflowing material and the mass
accretion rate $\dot{m}$ of the disk from which the material originates.  All trajectories shown have a starting
disk position of $r_{0}=100$ km and wind parameter $\beta=0.2$, which corresponds to rapid initial acceleration.
The shaded regions indicate mass fractions of $^{56}$Ni in excess of 0.5 (darkest shaded region), 0.4, 0.3, 0.2,
0.1, 0.05, and less than 0.05 (unshaded region).\label{fig:ni}}
\end{figure}

The production of nickel depends sensitively on the entropy per baryon in the outflowing material. 
Fig.~\ref{fig:ni} shows the final mass fraction of $^{56}$Ni in the disk outflow as a function of the mass
accretion rate $\dot{m}$ of the AD-BH and the entropy $s/k$ in the outflow, for a sample set of trajectories with
starting disk position $r_{0}=100$ km and wind parameter $\beta=0.2$.  We see vigorous production of $^{56}$Ni
over the majority of the parameter space.  Interestingly, for $\dot{m}\geq 0.2$, the production of nickel does not
appear to depend strongly on the accretion rate of the AD-BH and is instead a much stronger function of the
entropy per baryon in the outflow. 

\subsubsection{$\dot{m}\geq 0.2$}

First we consider the nickel production in outflows from the AD-BHs with $\dot{m}\geq 0.2$.  In the outflow
trajectories with $s/k\leq 20$, the entropy is not much higher than in the disk itself and so the material
retains some of the neutron richness of the disk.  In these cases, the neutron excess remaining after
alpha-particle assembly leads to neutron captures on the alpha-capture chain nuclei, and heavier isotopes of
nickel predominate.  Outflow trajectories with $s/k>20$ are sufficiently high in entropy such that the neutron
richness of the disk is erased by positron capture in the outflow.  The most efficient production of $^{56}$Ni
occurs just at this transition between neutron-rich and proton-rich, where $Y_{e}=0.5$; for the trajectories
considered in Fig.~\ref{fig:ni} this occurs when $s/k\sim 30$.  At even higher entropies, the assembly into
alphas is shifted to later times, and the weak reactions on free nucleons have more time to drive the material
proton rich.  Therefore as we consider outflow trajectories with various entropies, we find the higher entropy
cases produce higher $Y_{e}$ and thus less efficient production of $^{56}$Ni.

\subsubsection{$\dot{m}<0.2$}

The outflows from AD-BHs with $\dot{m}<0.2$ differ slightly from the above picture.  Disks with lower mass
accretion rates have lower central disk temperatures and densities.  The most important effect of this for disks
with $0.1\leq \dot{m}<0.2$ is that the outflows have lower starting temperatures and densities.  This means
the weak rates are not necessarily fast enough for the electron fraction to reequilibrate in the outflow prior to
nuclear reassembly, and so the neutron richness of the disk is retained to higher entropies.  For the lowest mass
accretion rate disks considered in this work, $0.05\leq \dot{m}<0.1$, the central disk temperatures and
densities are sufficiently low that the composition of the disk remains fairly balanced between neutrons and
protons, as shown in Fig.~\ref{fig:diskye}.  The outflows from these disks therefore have electron fractions
close to a half---and correspondingly strong $^{56}$Ni production---at lower entropies.

\subsection{Variation with $\beta$} \label{beta}

\begin{figure}[h]
\epsscale{.65}
\plotone{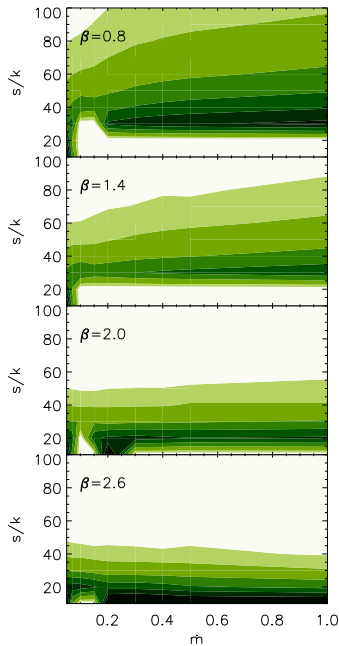}
\caption{Similar to Fig.~\ref{fig:ni}, but for wind parameters $\beta=0.8$, 1.4, 2.0, and 2.6.
\label{fig:beta}}
\end{figure}

While the entropy per baryon in the outflow plays a central role in setting the electron fraction, another important
ingredient is the time available for each stage of the nucleosynthesis.  In our outflow parameterization this is set
by the wind parameter $\beta$, which determines the initial acceleration of the outflow material. 
Fig.~\ref{fig:beta}, similarly to Fig.~\ref{fig:ni}, shows the final mass fraction of nickel in the outflowing
material as a function of the mass accretion rate $\dot{m}$ of the AD-BH and the entropy $s/k$ in the outflow, for a
sample set of trajectories with starting disk position $r_{0}=100$ km and wind parameters $\beta=0.8$, 1.4, 2.0, and
2.6.  As shown in Fig.~\ref{fig:beta}, increases in $\beta$ decrease the overall efficiency of $^{56}$Ni production
and, for high values of $\beta$, shift the entropy per baryon $s/k$ of the most efficient production. 

In order to understand the decrease in efficiency with increasing $\beta$, consider that the most robust $^{56}$Ni
production occurs when the outflow conditions evolve (1) sufficiently rapidly at early times to prevent the material
from becoming too proton-rich and (2) sufficiently slowly at late times to facilitate the complete conversion of
alphas to $^{56}$Ni.  Small values of $\beta$ correspond to rapid initial acceleration where the material quickly
attains coasting speed, while outflows with large values of $\beta$ accelerate slowly at first and evolve more
rapidly at later times.  This effect is illustrated in Table 1, which shows the average values of the outflow
velocity $\dot{r}$ divided by the radial distance of the material $r$ for two key points in the evolution of the
outflow---at early times when the electron fraction is set by weak interactions and late times during nuclear
reassembly.  The value of $\dot{r}/r$ serves as an indicator of how rapidly conditions evolve during these key
periods; robust $^{56}$Ni production therefore requires large values of $\dot{r}/r$ at early times and small values
at late times.  As Table 1 shows, these conditions are satisified in trajectories with the smallest values of
$\beta$.  Trajectories with larger $\beta$ evolve too rapidly at late times to efficiently synthesize
nickel, and, as shown in the table, most of the material is left as alpha particles.

\begin{figure}[h]
\epsscale{0.8}
\plotone{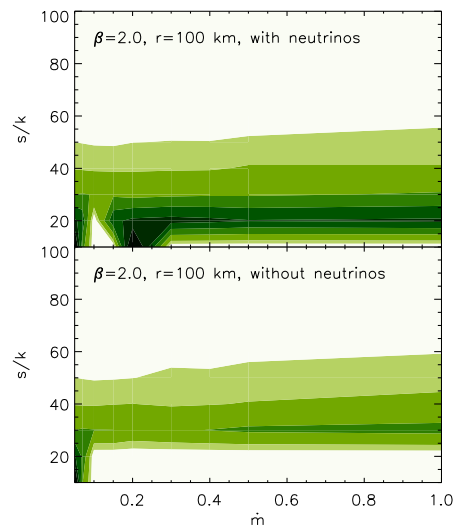}
\caption{The top panel is identical to the third panel of Fig.~\ref{fig:beta}, which shows nickel production as
in Fig.~\ref{fig:ni} for wind parameters $\beta=2.0$ and $r_{0}=100$ km.  The bottom panel is identical except
the neutrino interactions in the outflow have been turned off.
\label{fig:nu}}
\end{figure}

For very high $\beta$, $\beta\geq 2.0$, the most efficient $^{56}$Ni production shifts to lower entropies.  This
effect is due to the influence of neutrino interactions in the outflow.  As shown in Table 1, the evolution of
the outflow for high $\beta$ is slow at early times, when the material is close to the disk.  The high $\beta$
trajectories, therefore, receive the highest neutrino fluence.  Since the electron neutrino flux from the disk is
significantly higher than the electron antineutrino flux, the net influence of the neutrinos is to drive the
material proton rich via the reverse reaction in Eqn. 1.  The low entropy trajectories that, in the absence of
the neutrino interactions, would be neutron rich become proton rich and can therefore effectively produce
$^{56}$Ni, as illustrated in Fig.~\ref{fig:nu}.

\subsection{Variation with starting disk position} \label{diskr}

\begin{figure}[h]
\plotone{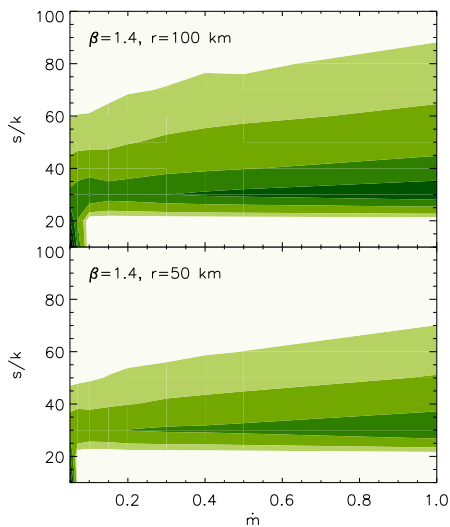}
\caption{Similar to Fig.~\ref{fig:ni}, but for wind parameter $\beta=1.4$ and starting disk positions $r_{0}=100$
km and 50 km.
\label{fig:r0}}
\end{figure}

While the outflow nucleosynthesis is primarily dependent on the outflow conditions as described above, the initial
conditions---the local disk conditions at the point of origin of the outflow---also have some bearing on the final
nickel production.  Fig.~\ref{fig:r0}, similarly to Fig.~\ref{fig:ni}, shows the mass fraction of nickel in the
outflowing material as a function of the mass accretion rate $\dot{m}$ of the AD-BH and the entropy $s/k$ in the
outflow, for a sample set of trajectories with $\beta=1.4$ and two starting disk positions, $r_{0}=50$ km and 100 km. 
Nickel production is similar in both cases, though trajectories originating from the inner disk region are somewhat
less efficient.  This is due to two separate effects---the influence of neutrino interactions and the starting
temperature of the material.  In the disk itself, the temperatures and densities increase with decreasing radius, and
correspondingly the rates of weak reactions, and therefore the emission of neutrinos, also increase.  Thus the
material outflowing from smaller radii will have higher initial temperatures and densities than outflows from larger
disk radii, and the neutrino flux the material is exposed to in the outflow will be higher.  The latter effect tends
to increase the electron fraction of the material; for trajectories that are already proton rich this decreases the
efficiency of alpha particle formation.  The former effect---the larger of the two---postpones alpha particle
reassembly to later times and farther distances in the outflow.  As shown in Table 1, the velocity divided by radial
distance $\dot{r}/r$ is faster during the epoch of alpha particle reassembly for smaller starting radii, resulting in
less efficient nickel production.

\section{Additional nucleosynthesis}

While in this work our focus is on the production of Nickel-56, we note that other interesting nuclear species may also
be synthesized in the outflow.  Trajectories with strong Nickel-56 production can make a number of intermediate mass
nuclei that are underproduced in galactic chemical evolution studies, such as $^{45}$Sc, $^{49}$Ti, and $^{64}$Zn
\citep{pru05}.  Low entropy ($s/k\leq 20$) trajectories from disks with $\dot{m}>0.1$ are moderately neutron rich; these
outflows produce heavier nickel isotopes and $N=50$ nuclei, and can potentially make some light $r$-process nuclei. 
High entropy ($s/k\geq 50$) trajectories with slow initial acceleration can make some interesting light $p$-process
nuclei such as $^{74}$Se, $^{78}$Kr, $^{84}$Sr, and $^{92}$Mo.

\begin{figure}[h]
\plotone{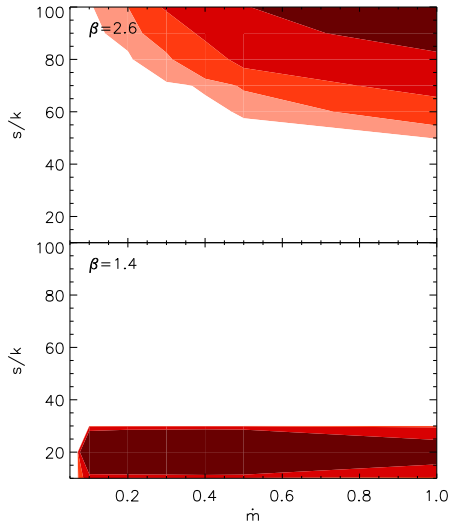}
\caption{Shows the final mass fraction of light $p$-process nucleus $^{92}$Mo divided by its solar value, for wind
parameters $\beta=1.4$ (bottom panel) and $\beta=2.6$ (top panel), and $r_{0}=100$ km. The shaded regions indicate
$^{92}$Mo production ratios in excess of 10000 (darkest shaded region), 1000, 100, and 10 (lightest shaded region).
\label{fig:Mo92}}
\end{figure}

An example of the latter two types of nucleosynthesis is shown in Fig.~\ref{fig:Mo92}, which compares the final mass
fraction of the light $p$-nucleus $^{92}$Mo produced in the outflow to its solar mass fraction.  The origin of this
nucleus has long been a mystery, though studies of the newly-discovered $\nu p$ process \citep{fro06,pru06} suggest
early supernova neutrino-driven winds to be a strong candidate site for its production.  Here we examine its production
in the context of our GRB outflow models. 

In the case of the moderately accelerating outflow, $\beta=1.4$, we see strong $^{92}$Mo production for low entropies,
where the material is moderately neutron-rich.  However, here the synthesis of $^{92}$Mo is accompanied by excessive
overproduction of some $N=50$ closed shell nuclei such as $^{88}$Sr, $^{89}$Y, and $^{90}$Zr.  The overproduction of
$N=50$ nuclei is so large (greater than a factor of $10^{5}$ over solar in some cases) for these trajectories that if
one knew precisely the amount of disk outflow and the rate at which accretion disks are formed, one could in principle
constrain the disk wind conditions from this information. 

Alternately, we see that for the case of the outflows with slow initial acceleration, $\beta=2.6$, the greatest
$^{92}$Mo production occurs at high entropies, from disks with the most vigorous neutrino emission.  These trajectories
have electron fractions greater than 0.6 and experience the highest neutrino fluence, both requirements for a successful
$\nu p$ process.  The production of $^{92}$Mo and other light $p$ nuclei via a $\nu p$ process in GRB disk outflows has
been discussed in \cite{kiz10}. 

\section{Alternate outflow model}

Finally, for comparison we consider an alternate, spherical outflow model in which the evolution of the electron fraction
is tied more closely to the hydrodynamics.  We construct the model by first constraining the outflow to follow radial
streamlines with zero vorticity.  This, along with the steady flow approximation, implies the Bernoulli function is
constant along the streamline that defines the outflow trajectory.  In its simplest form, the Bernoulli function is
\begin{equation}
B=\frac{1}{2}v^{2}+h+\Phi,
\end{equation}
where $v$ is the velocity, $h$ is the specific enthalpy, and $\Phi$ is the gravitational potential due to the black hole,
$-GM_{BH}/r$.  We write the specific enthalpy as
\begin{equation}
h=\frac{1}{\rho}\left[\sum_{i} n_{i}\mu_{i} + (kT)(s/k)n_{B}\right],
\end{equation}
where $n_{B}$ is the baryon number density, and $n_{i}$ and $\mu_{i}$ are the number density and chemical potential,
respectively, of species $i$, where the sum over species includes electrons, positrons, neutrons, protons, alpha
particles, and an average heavy nucleus.  We solve for the outflow hydrodynamics by further assuming adiabatic flow 
and a constant mass outflow rate, $\dot{M}=4\pi r^{2} \rho v$.  We evolve the composition of the outflowing material as
described in Sec. 2.

\begin{figure}[h]
\plotone{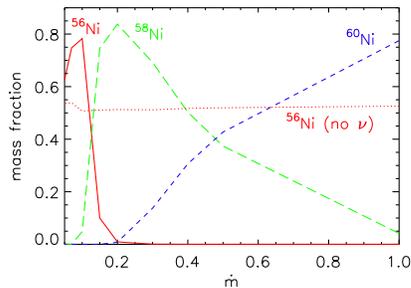}
\caption{Shows the final mass fractions of $^{56}$Ni (solid red line), $^{58}$Ni (long-dashed green line), and $^{60}$Ni (short-dashed blue
line) as a function of disk mass accretion rate $\dot{m}$, for the spherical outflow model with $s/k=40$ and
$r_{0}=50$ km, compared to the final mass fraction of $^{56}$Ni (dotted red line) obtained with the same outflow
model but without neutrino interactions.
\label{fig:spherical}}
\end{figure}

The spherical outflow model differs from the parameterized wind primarily in the shape of the outflow
trajectory---which is purely radial with no initial vertical piece---and the significantly slower acceleration of the
outflowing material.  Both of these effects impact the extent to which neutrino interactions influence the resulting
nucleosynthesis, though in opposite directions.  A purely radial trajectory moves material away from the central disk
region, which is the source of the hottest neutrinos, more directly than initially-vertical flow.  However, this
effect is dwarfed here by the effect of the dramatically larger neutrino fluence that arises due to the slower
acceleration.  The initial flow is so much slower that the electron fraction re-equilibrates in the outflow in every
case, to values above 0.5.  For the low mass accretion rate disks, this results in vigorous $^{56}$Ni production, as
shown in Fig.~\ref{fig:spherical}.  However, in higher accretion rate disks that have non-negligible antineutrino
emission, the production of nickel shifts to heavier isotopes.  This is due to a mini-$\nu p$-like process, whereby
antineutrino interactions on the free protons remaining after the onset of heavy element formation create neutrons
that are quickly captured by nuclei, primarily the dominant $^{56}$Ni.  The result is a shift from $^{56}$Ni to
$^{58}$Ni and finally to $^{60}$Ni with increasing disk $\dot{m}$, as shown in Fig.~\ref{fig:spherical}, compared to
the almost constant production of $^{56}$Ni that results when neutrino interactions are neglected.

\section{Conclusion}

Nickel production in the outflows from black hole accretion disks may be an important component of the total GRB Nickel-56
production.  We have outlined the ways in which $^{56}$Ni can be formed by black hole accretion disks under a variety of
conditions, taking careful account of the neutrino interactions. 

We find that entropy is a strong diagnostic across a broad range of accretion disk models.  Nickel production appears to be
optimal for only modest increases in entropy over the in-disk values; for the disks studied here with $\dot{m}>0.1$, this
can be accounted for via neutrino heating alone.  We note that nickel can also be produced via thermonuclear burning from a
shock that goes through a star, if the disk is associated with a hypernova explosion.  However, in models without this
additional production mechanism, one could constrain the required conditions for nickel production in the object.  For
example, if accretion disk outflows are to be a significant fraction of the overall GRB nickel production, the contribution
of other processes, such as magnetic heating, to the entropy increase in the outflow from $\dot{m}>0.1$ disks should not
greatly exceed that of the neutrino interactions.

However, conclusions about the mass fraction of Nickel produced cannot be drawn from the entropy alone.  Nickel formation
is optimized when the outflow timescale, in particular the ratio of $\dot{r}/r$ in the outflow, is fast at the beginning
stages of nucleosynthesis and slow later.  Furthermore, in the models considered here, outflows from an intermediate disk
radius, $r_{0}\sim 100$ km, more efficiently produced nickel than material ejected closer to the black hole.  Under the
optimal conditions up to around 50\% of the material ejected from black hole accretion disks will become Nickel-56.  These
results, when coupled with future wind model calculations and observations, can ultimately be used to constrain the outflow
conditions of GRB disks.

\acknowledgments
This work was partially supported by the Department of Energy under contract DE-FG02-02ER41216 (GCM) and under
contract DE-FG02-05ER41398 (RS), and the National Science Foundation ADVANCE Grant 0820032 (RS).

\begin{table}
\centering
\begin{tabular}{|c||c|c|c|c|}
\tableline & \multicolumn{4}{c||}{$r_{0}=100$ km}\\
\cline{2-5}
$\beta$ & $\dot{r}/r$, (a) & $\dot{r}/r$, (c) & $X_{^{56}\mathrm{Ni}}$ & $X_{\alpha}$ \\
\tableline \tableline
0.2 & 105  & 2.3 & 0.612 & 0.314 \\ 
0.8 & 32  & 6.1 & 0.488 & 0.449 \\
1.4 & 4.8 & 14 & 0.327 & 0.590 \\
2.0 & 0.62 & 28 & 0.189 & 0.667 \\
2.6 & 0.07 & 32 & 0.131 & 0.635 \\
\tableline & \multicolumn{4}{c||}{$r_{0}=50$ km}\\
\cline{2-5}
$\beta$ & $\dot{r}/r$, (a) & $\dot{r}/r$, (c) & $X_{^{56}Ni}$ & $X_{\alpha}$ \\
\tableline \tableline
0.2 & 120 & 3.1 & 0.582 & 0.355 \\
0.8 & 52 & 9.6 & 0.422 & 0.523 \\
1.4 & 5.2 & 27  & 0.229 & 0.689 \\
2.0 & 0.62 & 58 & 0.079 & 0.752 \\
2.6 & 0.06 & 64 & 0.048 & 0.705 \\
\tableline
\end{tabular}
\caption{Average values of velocity divided by radial distance $\dot{r}/r$ in units of s$^{-1}$ for two sets of
trajectories with entropy $s/k=30$ and starting disk positions $r_{0}=100$ km and $r_{0}=50$ km from an AD-BH
with $\dot{m}=0.5$. Each set of trajectories includes five values of wind parameter $\beta$, and the average
$\dot{r}/r$ is calculated twice: (a) at early times as the electron fraction is set ($T_{9}>10$) and (c) at
later times during nuclear reassembly ($6>T_{9}>2$), where (a) and (c) correspond to the portions of the
trajectory depicted in Fig.~\ref{fig:traj}.  Also included are the final mass fractions of $^{56}$Ni and alpha
particles for each trajectory.\label{table:taueff}}
\end{table}


\begin{thebibliography}{}
\bibitem[Bloom et al.(1999)]{bloom99} Bloom, J.~S., et al.\ 
1999, \nat, 401, 453 
\bibitem[Chen \& Beloborodov(2007)]{che07} Chen, W.-X. \& Beloborodov, A., 2007, \apj, 657, 383
\bibitem[Frohlich et al.(2006)]{fro06} Frohlich, C. et al. 2006, \apj, 637, 415
\bibitem[Fryer et al.(2006)]{fryer06} Fryer, C.~L., Young, 
P.~A., \& Hungerford, A.~L.\ 2006, \apj, 650, 1028 
\bibitem[Fujimoto et al.(2003)]{fuji03} Fujimoto, S., Hashimoto, M., Koike, 
Arai, K., \& Matsuba, R. 2003 \apj, 585, 418 
\bibitem[Galama et al.(1998)]{galama98}
Galama, T.~J., et al. 1998, Nature, 395, 670
\bibitem[Hjorth et al.(2003)]{hjorth03} Hjorth, J., et al.\ 
2003, \nat, 423, 847 
\bibitem[Kawabata et al.(2003)]{kawabata03}
Kawabata, K. et al. 2003, ApJ, 593, L19
\bibitem[Kizivat et al.(2010)]{kiz10}
Kizivat, L.-T., Martinez-Pinedo, G., Langanke, K., Surman, R., \& McLaughlin, G.~C. 2010,
\prc, 81, 025802
\bibitem[MacFadyen \& Woosley(1999)]{macfadyen99} MacFadyen, A. I. \& 
    Woosley, S. E. 1999, \apj, 524, 262
\bibitem[Maeda \& Nomoto(2003)]{maeda03} Maeda, K.~\& Nomoto, 
K.\ 2003, \apj, 598, 1163 
\bibitem[Maeda 
\& Tominaga(2009)]{maeda09} Maeda, K., \& Tominaga, N.\ 2009, \mnras, 288 
\bibitem[Malesani et al.(2004)]{malesani04}
Malesani D., et al. 2004, \apj, 609, L5
\bibitem[Matheson et al.(2003)]{matheson03}
Matheson T., et al. 2003, \apj, 599, 394 
\bibitem[Pruet, Woosley, \& Hoffman(2003)]{pru03} Pruet, J., Woosley, 
    S. E., \& Hoffman, R. D. 2003, \apj, 586, 1254
\bibitem[Pruet, Thompson, \& Hoffman(2004)]{pru04} Pruet, J., Thompson, T.,
\& Hoffman, R.~D. 2004, \apj, 606, 1006
\bibitem[Pruet et al.(2006)]{pru06} Pruet, J., Hoffman, R.~D., Woosley, S.~E., Janka, H.-T., \& Buras, R. 2006, 
\apj, 644, 1028
\bibitem[Pruet, Surman, \& McLaughlin(2005)]{pru05} Pruet, J., Surman, R.,
\& McLaughlin, G.~C. 2005, \apj, 602, L101
\bibitem[Qian \& Woosley(1996)]{qia96} Qian, Y.-Z. \& Woosley, S.~E. 1996, \apj, 471, 331
\bibitem[Stanek et al.(2003)]{stanek03} Stanek, K.~Z., et al.\ 
2003, \apjl, 591, L17 
\bibitem[Surman \& McLaughlin(2004)]{sur04} Surman, R.~\& 
McLaughlin, G.~C. 2004, \apj, 603, 611
\bibitem[Surman \& McLaughlin(2005)]{sur05}
  Surman, R. \&  McLaughlin, G. C. 2005, \apj, 618, 397 
\bibitem[Surman et al.(2006)]{sur06} Surman, R., McLaughlin, 
G.~C., \& Hix, W.~R.\ 2006, \apj, 643, 1057 
\bibitem[Thomsen et al.(2004)]{thomsen04}
Thomsen B., et al. 2004, A\&A, 419, L21 
\bibitem[Woosley(1993)]{woosley93} Woosley, S. E. 1993, \apj, 405, 273
\bibitem[Woosley \& Bloom(2006)]{woosley06} Woosley S.E., Bloom J.S. 2006, ARAA, 44, 507 
\bibitem[Woosley and Heger(2003)]{woosley03} Woosley, S. E. and Heger, A. 2003, preprint (astro-ph/0309165)
\end{thebibliography}
\end{document}